\documentclass[proof]{WileyASNA-v1}

\usepackage[T1]{fontenc}
\usepackage[utf8x]{inputenc}
\usepackage[portuguese,english]{babel}

\articletype{Article}%

\received{}
\revised{}
\accepted{}

\begin{document}

\title{Spiral structure of the galactic disk and its influence on the rotational velocity curve%\protect\thanks{This is an example for title footnote.}
}

%\author[1]{Author One*}

%\author[2,3]{Author Two}

%\author[3]{Author Three}

%\authormark{AUTHOR ONE \textsc{et al}}

%\address[1]{\orgdiv{Org Division}, \orgname{Org Name}, \orgaddress{\state{State name}, \country{Country name}}}

%\address[2]{\orgdiv{Org Division}, \orgname{Org Name}, \orgaddress{\state{State name}, \country{Country name}}}

%\address[3]{\orgdiv{Org Division}, \orgname{Org Name}, \orgaddress{\state{State name}, \country{Country name}}}

%\corres{*Corresponding author name, This is sample corresponding address. \email{authorone@gmail.com}}

%\presentaddress{This is sample for present address text this is sample for present address text}

\author[1]{Miroslava Vukcevic*}
\author[2,3]{Vladimir Zekovic}
\author[2,4]{Marko Radeta}
\authormark{Vukcevic \textsc{et al}}

\address[1]{\orgname{Astronomical Observatory}, \orgaddress{Volgina 7, 11000 Belgrade, Serbia}}
\address[2]{\orgname{Department of Astronomy, Faculty of Mathematics, University of Belgrade}, \orgaddress{Studentski trg 16, 11000 Belgrade, Serbia}}
\address[3]{\orgname{Department of Astrophysical Sciences, Princeton University}, \orgaddress{Princeton, NJ 08544, USA}}
\address[4]{\orgname{Wave Labs, MARE - Marine and Environmental Sciences Centre, Agência Regional para o Desenvolvimento da Investigação, Tecnologia e Inovação (ARDITI), University of Madeira}, \orgaddress{Polo Cientifico e Tecnologico da Madeira, 9020-105 Funchal, Madeira, Portugal}}

\corres{*Miroslava Vukcevic, \email{vuk.mira@gmail.com}}
\presentaddress{Astronomical Observatory, Volgina 7, 1100 Belgrade, Serbia}

\abstract{The most spiral galaxies have a flat rotational velocity curve, according to the different observational techniques used in several wavelengths domain. In this work, we show that non-linear terms are able to balance the dispersive effect of the wave, thus reviving the observed rotational curve profiles without inclusion of any other but baryonic matter concentrated in the bulge and disk. In order to prove that the considered model is able to restore a flat rotational curve, Milky Way has been chosen as the best mapped galaxy to apply on. Using the gravitational N-body simulations with up to $10^7$ particles, we test this dynamical model in the case of  the Milky Way with two different approaches. Within the direct approach, as an input condition in the simulation runs we set the spiral surface density distribution which is previously obtained as an explicit solution to non-linear Schr\"{o}dinger equation (instead of a widely used exponential disk approximation). In the evolutionary approach, we initialize the runs with different initial mass and rotational velocity distributions, in order to capture the natural formation of spiral arms, and to determine their role in the disk evolution. In both cases we are able to reproduce the stable and non-expanding disk structures at the simulation end times of $\sim10^9$ years, with no halo inclusion. Although the given model doesn't take into account the velocity dispersion of stars and finite disk thickness, the results presented here still imply that non-linear effects can significantly alter the amount of dark matter which is required to keep the galactic disk in stable configuration.}


\maketitle

\section{Introduction}\label{sec1}

The rotation curve of spiral galaxies is the mean circular velocity around the core as a function of radius, measured by spectroscopic observations of emission lines such as $H_{\alpha}$, HI and CO lines from disk components, stars and gas. In the most  of observed curves the shape is almost flat implying the lack of mass in the outer part of galaxies. 
There are a number of articles and reviews on rotation curves and mass determination of galaxies (Sofue and Rubin 2001). In most of them lack of mass has been overcome by introducing the spherical halo surrounding the galactic bulge and disk. So that, in the galactic dynamics investigation there are three separate mass distributions: bulge, disk and halo. 
Importance of the mass distribution rather than the mass value itself has already been pointed in the literature. The best example is given in Binney and Tremaine (1987), Fig. 2.17.

The disk mass is best approximated by an exponential distribution so far, meaning that there are equipotential rings with the exponentially decreasing mass distribution going further from the galactic center. Using this approximation, the mass trapped within spirals has been completely ignored. Although there is just 5-10 percent of the disk mass per spiral arm, it is very important to include this particular distribution into rotational velocity calculation. Here, we underline that considering the motion of the stellar component within the arms as noncircular, is not true along the whole spiral. Stars inside the bulge and in the very inner part of the disk, discard significantly from the circular motion, where our model breaks down. We will show that our model is valid from 1kpc up to the end of the disk, and at that radius, the domain pitch angle (an angle between the tangent on the spiral and tangent on the circle at given radius) is very small, being the order of few degrees (Binney and Tremaine 1987). So that, recent result observed by THINGS (Trachtenach et al. 2008) agrees with our nonlinear density wave model. Argument for the direction of the group velocity of the nonlinear wave has been explained in details in Vukcevic (2014).
There are other approaches trying to explain the observations with no additional mass inclusion: Modified Newtonian Dynamics (MOND) 
%\citet{Milgrom1983, Moffat2006, Carignan2013}
(Milgrom 1983; Moffat 2006; Carignan 2013), modifying the acceleration; modification of gravitational potential different from classical expression %\citet{Carignan2013, Mannheim2012}
(Carignan 2013; Mannheim 2012); numerically modeling for computing the Newtonian dynamics of thin-disk galactic rotation introducing new parameter called the galactic rotation number (Feng and Gallo 2011).  

The motivation to use nonlinear density perturbation solution in order to derive the rotation velocity expression is as follows: if for the dynamical system exists an integrable nonlinear wave equation, then the dynamics of that system is represented by a long lasting stable wave, with constant amplitude and group velocity. In the case of spiral galaxies such a solution has already been derived as one-dimensional wave curved due to the rotation, and following the spiral shape, as it is observed not only in optical but in other wavelengths, as well. Physically, it means that the density is enhanced along the spiral, due to the gravitational potential which traps stars and gas.
Therefore, the velocity of the mass trapped by the wave is enhanced along the formed wave, and it can explain the flat, constant rotational curve. It will remain stable as long as the conditions are satisfied, namely as long as the disk is at certain stability regime, approximately at least one rotation period 
%\citet{Vukcevic2014}
(Vukcevic 2014). The rotation velocity curve derived using this theory, follows the observed curve due to the constant epicyclic frequency to surface mass density (SMD) ratio (although these two parameters are radius dependent). The parameters involved in the velocity expression are consequence of the nonlinear density wave solution, and are to be derived directly from the observational data, and not by any fitting procedure.

The soliton model for spiral arms has the following advantages: it overcomes the main difficulty of the maintaining mechanism; there is a fine structure inside the soliton with space period much smaller than the width of the soliton, possibly explaining the star formation process; as the soliton emerges at the edge of the disk it transports the material into outer regions, keeping the disk at the threshold of instability; the last prediction can interpret distribution of neutral gas clouds at the periphery of the galaxy.
We have chosen our host galaxy Milky Way to apply the model on, and to derive the rotational velocity curve from the theory.
The model has been tested by N-body simulations performed for the Milky Way in two different approaches: the first, direct approach, applies the nonlinear spiral density soliton for the surface mass density of the disk (instead of exponential one), and the second, evolutionary approach, uses the nonlinear vortex soliton solution (Vukcevic 2019) for the initial mass distribution. 

Appart of the disk galaxies, these soliton solutions can be applied in some other astrophysical systems, such as accretion disks or planetary atmospheres.

\section{Method}\label{sec2}

In this paper, a theoretically derived expression for the surface mass density profile in weak non-linear regime, is used to obtain the rotational velocity curve. The stellar component of the galaxy is described by standard fluid equations, together with Poisson's equation. The model is considered as an infinitesimally thin and cold disk.

The dynamical balance between the gravitational and centripetal forces is used to calculate the rotational velocity $V$ of the test particle:

\begin{equation}
	V^{2}(r)=\frac{GM(r)}{r}=r\frac{\partial \phi}{\partial r},
\end{equation}
where $G$ is gravitational constant, and $M$ is total mass producing the gravitational force acting on the test particle, and consequently gravitational potential. The main task is to determine as best as possible the gravitational potential dependence on the radius.

Using the expansion of variables done by Vukcevic 
%\citep{Vukcevic2014}
(Vukcevic 2014), the gravity potential gradient $\partial \phi/\partial r$ is approximated by

\begin{equation}
	\frac{\partial \phi}{\partial r}=r\Omega^{2}+\sum^{n=1}_{\infty}\sum^{\infty}_{m=-\infty} 2\pi G \epsilon^{n}\Re(\rho^{(n,m)}(\xi,\eta)e^{i(kr-\omega\tau)}),
	\end{equation}
where term $r\Omega^{2}$ comes out from the equilibrium property. $\Omega$ is angular velocity, $G$ is gravitational constant, $\rho$ denotes surface mass density, $\xi$, $\eta$ and $\tau$ are corresponding stretched coordinates, while $k$ and $\omega$ are wave number and frequency, respectively.

In order to derive the gravitational potential by which stars are trapped, we have used already derived solution to Nonlinear Schr\"odinger Equation for the surface density perturbation:

	\begin{equation}
	\rho^{1,1}(\xi,\eta)=\rho_{a}\frac{e^{i\psi}}{ch(\sqrt{\frac{Q}{P}}\rho_{a}(\xi-P\eta))}.
		\end{equation}
		
The wave phase $\psi$ is not relevant for rotational velocity since only the real part of Equation (3) is taken, and parameters $P=\frac{1}{2}\kappa/\pi G\rho_{0}=1/V_{g}$ and $Q=\kappa^{3}/\pi G\rho_{0}$ are related to the soliton velocity $V_{g}$, and to the soliton width; $\rho_{a}$ is wave amplitude, $\kappa$ is epicyclic frequency value,  $T=1=(t+\varphi/\Omega)$, where $\varphi$ is polar angle.

Substituting the exact solution for the density given in Equation (3), into the potential gradient in Equation (2), the expression for the rotational velocity given in Equation (1) reads as

\begin{equation}
	V(r)=\sqrt{\Omega^{ 2}r^{2}+\frac{ar}{ \cosh b(T-cr)}}.
	\end{equation}

All parameters and variables are dimensionless. By returning to the original coordinates (multiplying the non-dimensional velocity by $2\pi G \rho_{0}/\kappa$, and $T$ by $\kappa$) and by taking into account that the time is evaluated as $10^{7}$ yr, while $r$ is given in $[kpc]$ and $V$ in $[km s^{-1}]$, it is possible to derive the parameters $a$, $b$, and $c$ in Equation (4).

\subsection{Derivation of parameters}

The parameter $a$, which is related to the amplitude of the wave $\rho_{a}$ (the density enhancement along the spiral), given as a number that accounts for $r$ expressed in $[kpc]$ and rotational velocity expressed in [$km s^{-1}$], reads as follows:  

\begin{equation}
a=2\pi G \rho_{0}\rho_{a} (3\times10^{16}) [km s^{-2}].
\end{equation}

The parameter $b$ is the one related to the relative wave amplitude and epicyclic frequency $\kappa$, and it reads as follows:
\begin{equation}
b=\frac{1}{2}\kappa \rho_{a} (3\times10^{16})[s^{-1}], 
\end{equation}
while parameter $c$ is a constant which is related to the soliton velocity

\begin{equation}
c=\frac{1}{V_{g}} [s km^{-1}] 
\end{equation}
since $c$ is multiplied by $r$ expressed in $[kpc]$. For typical values ($\kappa\sim10^{-15}s^{-1}$, $\rho_{0}\sim (4-6)\times10^{-2} gcm^{-2}=(200-300)M_{\odot} pc^{-2}$, and $\rho_{a} \sim 0.4$), the group velocity is approximately $V_{g}=2\pi G \rho_{0}/\kappa\sim200kms^{-1}$. This wave velocity coincides with the rotational velocity of particles as long as the soliton wave exists, since the group velocity is tangent to the spiral at given $r$, while rotational velocity is tangent to the circle at that same $r$. Therefore, $V_{g}=V\cos \alpha$ where $\alpha$ is an angle between the tangent on the spiral and tangent on the circle at given $r$. This angle is very small, and $\cos \alpha\simeq 1$. The direction of the group velocity represents the direct consequence of the coordinate transform procedure in the derivation of non-linear equation due to marginal stability regime of spiral galaxies 
%\citet{Bertin2000}
(Bertin 2000). In that case, the frequency of the wave goes to zero, while the group velocity tends to infinity, meaning that the special treatment of the coordinates transform is required 
%\citet{Watanabe1969}
(Watanabe 1969).

%%%%%ovde moras da objasnis da ovo nisu free parametri koje fitujes da bi ti kriva ispala kako hoces vec ih dobijas iz posmatranja i pri tom nisu nezavisna!!!!!  Takodje da ubacis da exp disk daje decreasing bergen ili sofue ili binney tremaine

%We expect theoretical rotational curve to be valid from $r\sim1$kpc ($\kappa\sim10^{-15} s^{-1}$, $\rho_{0}\sim (4-6)\times10^{-2} g cm^{-2}=(200-300)M_{\odot} pc^{-2}$) and density enhancement about 5-10 percent ($\rho_{a} \sim (0.4-0.8)$), up to infinity, theoretically, due to the linear dispersion analysis 
%\citet{Vukcevic2014}
%(Vukcevic 2014).

In the case of a flat rotation curve, the value of epicyclic frequency $\kappa$ can be approximated by $\sqrt{2}\Omega$ 
%\citet{Binney1987}
(Binney \& Tremaine 1987). Therefore, we are left with the three relevant parameters only -- the surface mass density, the wave amplitude that is normalized by SMD, and either $\Omega$ or $\kappa$.

We have not made an a priori assumption on surface density distribution, nor on gravity potential. The system has been treated self-consistently, and relevant parameters to be estimated using the observed data are: equilibrium surface density of the galaxy $\rho_{0}$, amplitude of the wave (density enhancement along  the spiral), and angular velocity. Since the three parameters could be estimated from the observational data for the given galaxy, it means that the derived equation for rotational velocity, Equation (4), is not a parametric one.

The shape of the rotational velocity solution is sensitive to the mentioned parameters. Since estimated values for angular velocity, surface stellar mass density, and relative amplitude of the surface density are method dependent, some of these values can differ in an order of magnitude for the same galaxy. In the first approximation, we have substituted the averaged values for $\rho_{0}$ and $\kappa$ (although they are both $r$ dependent) in order to show that Equation (4) is rather universal, and it can support the rotational curves of many spiral galaxies. 

The basic property of a soliton wave is the constant group velocity of the wave, as well as, the constant wave amplitude along the spiral. This particular property motivated us to expect a possible support to the observed rotational velocity curve without inclusion of any other but baryonic matter. If one thinks of the orbital velocity of a star at some certain distance from the center, then the star would be trapped by the potential due to the wave that passes by that radius. The wave will thus enlarge the velocity of the star. Since the wave is soliton, its group velocity will be constant. Therefore, the expression of rotational velocity is theoretically derived by keeping the nonlinear terms in the calculation. It is not derived by any kind of fitting process involved, neither in the derivation of expression, nor in the parameters derivation. The parameters are explicitly calculated using the SMD distribution function employed in the radial gradient of gravitational potential in order to derive rotational velocity. The SMD function is not assumed but rather derived as a solution of integrable nonlinear equation. All previously mentioned details provide us with a general expression that can be used for any spiral galaxy if it is possible to estimate the SMD and rotational velocity, for example. %Since the SMD can be derived from the rotational velocity curve or using mass to light ratio, we suggest to use values obtained by independent mass to light ratio method, while for estimation of angular velocity it has to be done by using the rotational velocity curve. 
It would be the best, if it is possible to estimate both values by using some method that is independent of the rotational velocity curve, as it was done for the Milky Way 
%\citet{Vukcevic2021}
(Vukcevic 2021).

Regarding the observed level of non-axisymmetric photometric and kinematic structure of spiral galaxies, our proposed model is compatible with the data presented in the Tranchternach et al. (Trachtenach et al. 2008), for instance. They have shown that non-circular motions are typically 4.5 percent on average of the mean circular velocity in a sample of about 20 spiral galaxies. In the proposed nonlinear density wave model, for the radius greater than 1kpc, the group velocity of stars trapped within the spiral is almost the same as the rotation or circular velocity of the stars at that same radius (Vukcevic 2021). The group velocity direction is defined by the expansion procedure of used Reductive perturbation method in order to derive an integrable nonlinear equation; it has the direction of a tangent on the spiral at the given radius. The circular or rotational velocity direction is the tangent on the circle at given radius. In each point of the disk it is possible to define a pitch angle (see Binney \& Tremaine) which theoretically is the order of $0.14^{0}$. It means that the discrepancy between the group and circular velocities in the disk is negligible. The observed value of the pitch angle is typically $14^{0}-15^{0}$ and this deviation between the observed and theoretical values is stated as a winding dilemma. The nonlinear spiral wave can persists for at least 2 orbital periods, contrary to the linear density waves, and can be taken as a transient equilibrium state.

\section{Results} 

Luminosity is one of the most important characteristics of a galaxy, as well as the rotation. It has been shown that the surface mass density distribution and the angular velocity function are not independent of each other (Vukcevic 2021). We have focused on the Milky Way and estimated the parameters $a$, $b$, and $c$ by using the values of surface mass density $\rho_{0}$ and $\kappa$ from results of Sofue (2018) and Fich (1989), respectively. Finally, we have applied it in Equation (4), thus obtaining the rotational velocity curve. 

The mass distribution and epicyclic frequency are both well mapped in the case of Milky Way, although there are a lot of difficulties in measurements due to our location within the galaxy. We base our derivation of rotation velocity curve on the data given by Luna et al. 
%\citep{Luna2006}
(2006). The curve is presented in Fig. 1 by dotted line, plotted for following values: surface mass density $\rho_{0}\sim 8\times10^{-2} gcm^{-2}=400M_{\odot}pc^{-2}$ 
%\citet{Sofue2018}
(Sofue 2018), $\rho_{a} \sim 0.4$, angular velocity $\Omega\sim 20kms^{-1} kpc^{-1}$ %\citet{Fich1989}
(Fich 1989) and consequently $\kappa\sim 0.9\times10^{-15}s^{-1}$. Therefore, we estimate the parameters as follows: $a\sim 8\times10^{3}kms^{-2}$, $b\sim 0.038 s^{-1}$, and $c\sim 3.9 skm^{-1}$. At the same figure, by solid line, we plot the rotational velocity curve of the Milky Way galaxy obtained from Eq. (4) by using appropriate radial functions for $\rho_{0}$ and $\Omega$, instead of the averaged values (Vukcevic 2021). The averaging procedure makes sense only for larger radii, where both functions slowly decline with radius.

\begin{figure}[h]
%\centering
\includegraphics[width=1.0\columnwidth]{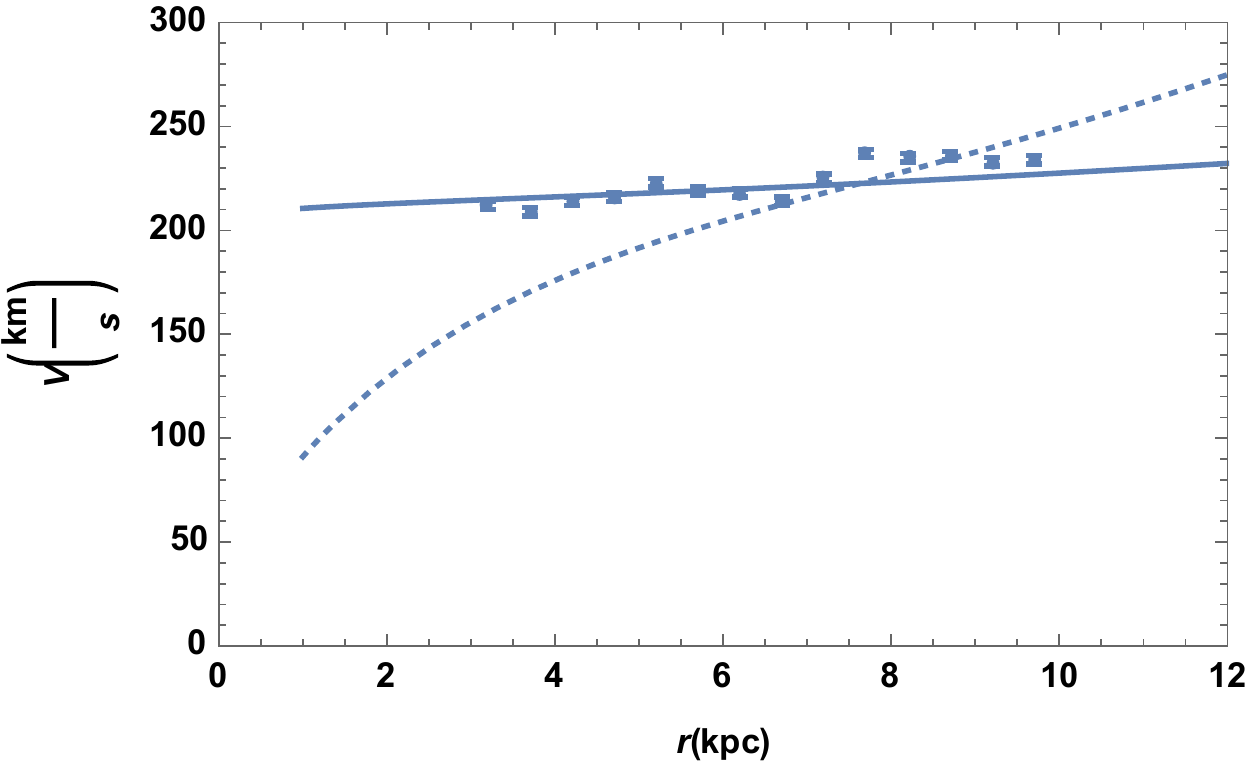}
\caption{The rotational velocity curve of the Milky Way galaxy. Dots represent the observed data (Luna 2006), dotted curve is the result of Equation (4) for following parameters: $a=8\times10^{3}$, $b=0.038$, $c=3.9$, and solid line is the result of Equation (4) for power-law functions of surface mass density and angular velocity (Vukcevic 2021).}
\label{fig1}
\end{figure}

\section{Probing the theory with gravitational N-body simulations
}\label{sec5}

The gravitational N-body simulations enabled us to probe the non-linear soliton solution on the Milky Way type of galaxy. We used the GADGET 2 code (Springel 2005) to run the 2D gravitational N-body simulations that consider only the star component by two different approaches. We did not include the dark matter (DM) halo in the runs.

\begin{figure*}[h!]
\centering
\includegraphics[width=0.33\linewidth]{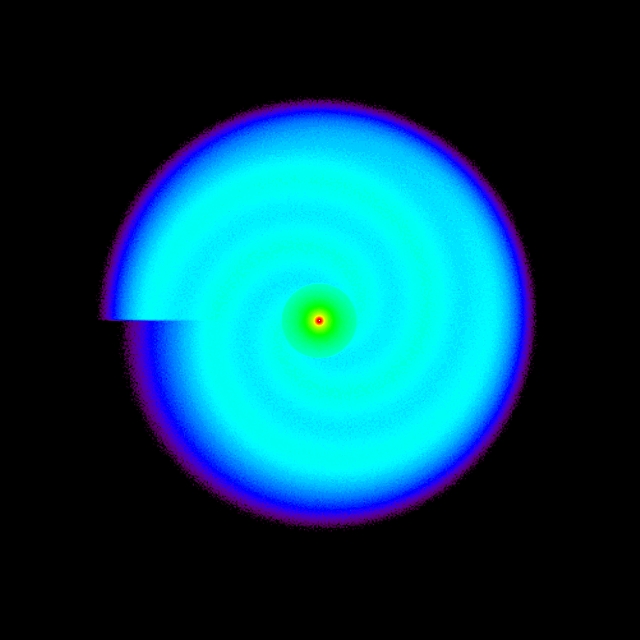}
\includegraphics[width=0.33\linewidth]{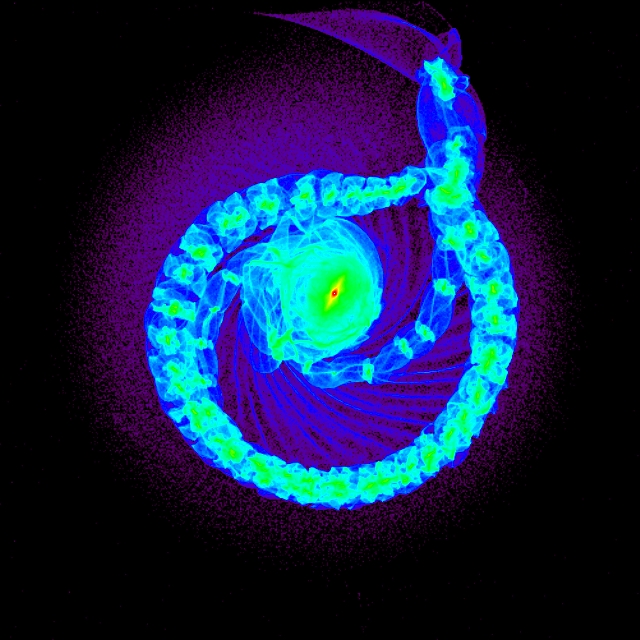}
\includegraphics[width=0.33\linewidth]{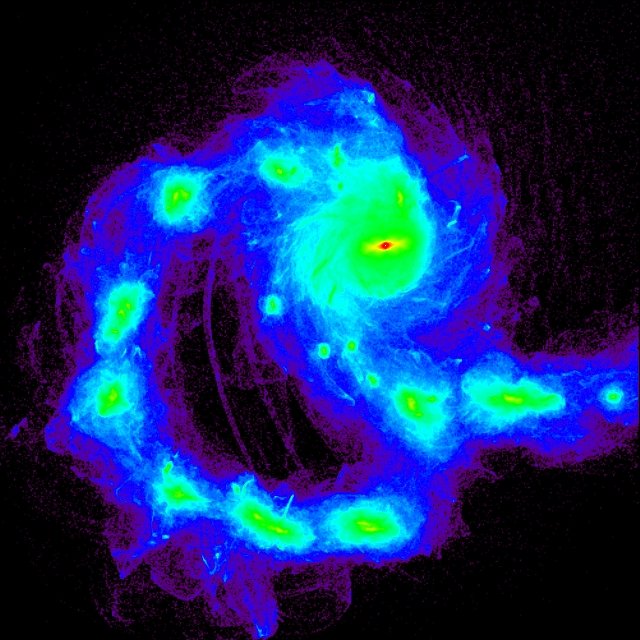}
\caption{The simulation run with the initial density distribution set as solution to the non-linear Schr\"odinger equation. The initial conditions are:
$M = 2\times 10^{11} M_{\rm sun}$, $N = 1.5 \times10^{7}$ particles, $V \sim 200 km/s$, $R \sim30 kpc$. The plots size is 100x100 $kpc$. The snapshot times are: 0, 0.25, and 0.5 $Gyr$ from left to right.}
\label{fig2}
\end{figure*}

\begin{figure*}[h!]
\centering
\includegraphics[width=0.33\linewidth]{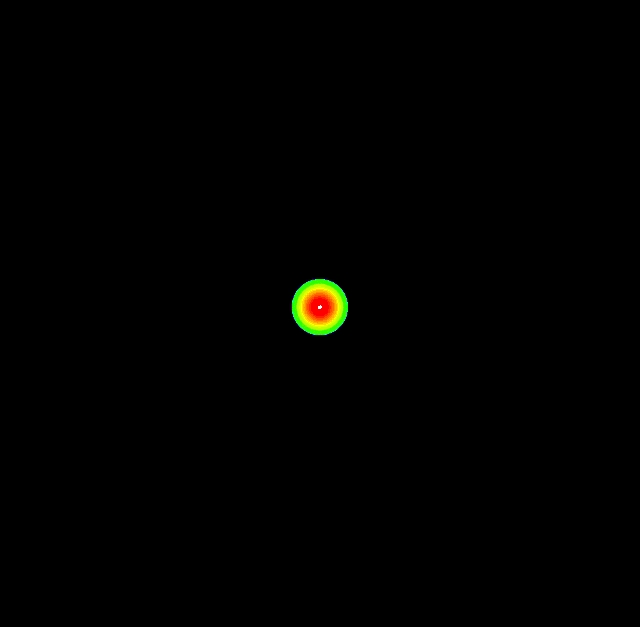}
\includegraphics[width=0.33\linewidth]{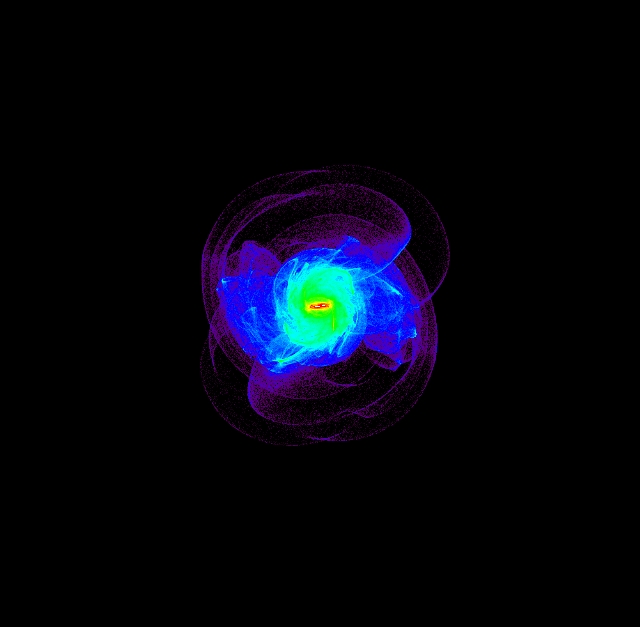}
\includegraphics[width=0.33\linewidth]{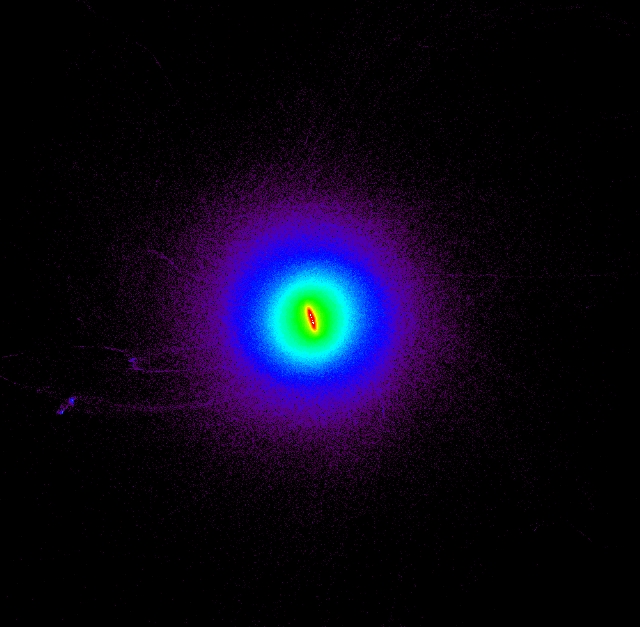}
\caption{The simulation of evolving initial mass, set as non-linear vortex solution. The initial conditions are: $M = 10^{11} M_{\rm sun}$, $N = 10^{7}$ particles, $R \sim 5 kpc$, $V \sim 200 km/s$. The plots size is 100x100 $kpc$. The snapshot times are: 0, 0.15, and 1 $Gyr$, from left to right.}
\label{fig3}
\end{figure*}

In the direct approach (see Fig. 2) as an input conditions we set the spiral surface density distribution and the velocity from Equations (3,4) for the disk, and axisymmetric $\rho \sim 1/(1+5 r)$ distribution for the bulge. As a result, we get that after the period of 1 Gyr the diffusing mass (with R > 50kpc) is less than 12 percent.

In the evolutionary approach (see Figure 3) we use an axisymmetric non-linear vortex solution for the initial mass density distribution. The expansion of the disk becomes stabilized over time, by the growth of spiral arms. The soliton structure (spiral arms) grows within the initial mass, which evolves into a stable (on the scale of $2\ Gyr$) disk+bulge configuration with $R \sim 40\ kpc$, with an outflowing mass of less than few percents at the simulation end time.

\section{Conclusions}\label{sec6}

The luminosity itself is not enough to derive a clear conclusion on the type of rotation curve one could expect. It is rather necessary to follow the ratio of two variables: the surface mass density, and, the angular velocity. 
%4 examples of spiral galaxies presented in this paper, show that it is possible to restore the rotational velocity curve of spiral galaxies keeping nonlinear terms in the variables expansion in order to derive disk surface mass density, with no inclusion of any other but baryonic matter. Examples are chosen as the best to estimate necessary physical values such as SMD and angular velocity. 
 
It is also necessary to test the model on a larger sample of galaxies, which is the subject of our further research. However, we can circumspectly expect that the rotational velocity expression which is derived with respect to nonlinear effects, in the case of spiral galaxies is a rather general one. Even though it has been derived under the infinitesimally thin disk approximation, it can be taken as an accurate one, due to the gas contribution that has been neglected.  
The obtained result is important from few points of view: first, the SMD function is analytically derived instead of being assumed (as e.g. exponential); second, by doing a parametric study on the proposed model within N-body simulations, can give a better insight into the physical inputs used in the simulations of galaxy dynamics; third, this stable nonlinear solution gives an opportunity to investigate the dynamics of the merging process of two or more galaxies throughout the soliton interaction; and fourth, this solution could eventually be compared to MOND theory in order to try to find out a possible physical explanation for the parameters introduced in MOND.

The preliminary simulation results suggest that the spiral distribution of our soliton model can support stable disk configurations for at least one $Gyr$, with no dark matter inclusion.

Finally, even the proposed method cannot cover all the issues regarding the missing mass in galaxies. It is however important in the reinvestigation of the minimal amount of dark matter that is required to overcome the difficulties in the rotation curve shape. This amount could be reduced, due to nonlinear effects in the disk dynamics.
However, this result is limited to the small scales only, and it does not resolve the problem of dark matter inclusion in clusters of galaxies. This large scale phenomena could be possibly reinvestigated in the nonlinear regime by using a method of multiple solitary wave interactions within the soliton perturbations.\\

Further research is devoted to the confirmation of the model on a larger sample of spiral galaxies with flat rotational curves.

%\pagebreak
%\backmatter

\section*{Acknowledgments}

During the work on this paper MV and VZ were financially supported by the Ministry of Education, Science and Technological Development of the Republic of Serbia through the contract No. 451-03-9/2021-14/200002 (MV) and No. 451-03-9/2021-14/200104 (VZ). The part of N-body simulations was run on the cluster SUPERAST at the Department of Astronomy, Faculty of Mathematics, University of Belgrade.

\subsection*{Author contributions}

MV derived the analytical soliton solution, while VZ and MR ran the N-body simulations in this work.

%\section*{Supporting information}

%\appendix
\section*{References}

%Babcock H. W. 1939, Lick Observatory Bulletin 19, p.41.\\
Bertin G. 2000, Dynamics of Galaxies, Cambridge Univ. Press, Cambridge.\\
Binney J., Tremaine S. 1987, Galactic Dynamics, Princeton Univ. Press, Princeton, NJ\\
%de Blok W. J. G., Walter F., Brinks E., Trachternach C., Oh S-H. \& Kennicutt Jr. R. C. 2008, The Astronomical Journal, 136, p.2648\\
%\bibitem{cardone12}
%Cardone V. F., Capone M., Radicella N. \& Ruggiero M. L. 2012, Mon. Not. R. Astron. Soc. 423, p.141\\
%\bibitem{carignan13} 
Carignan C., Frank B. S., Hess K. M., Lucero D. M., Randriamampandry T. H., Goedhart S. \& Passmoor S. S. 2013,  Astronomical Journal 146, p.48\\
%\bibitem[\protect\citeauthoryear{Carignan}{2006}]{Carignan2006}
%Carignan C., Chemin L., Huchtmeier  W. K., Lockman F. J. 2006, Astrophysical Journal, 641, L109
%\bibitem[\protect\citeauthoryear{Carignan}{2013}]{Carignan2013}
%Carignan C., Frank B. S., Hess K. M., Lucero D. M., Randriamampandry T. H., Goedhart S., Passmoor S. S. 2013, Astronomical Journal, 146, 48
%\bibitem{corbelli10} 
%Corbelli E., Lorenzoni S., Walterbos R., Braun R. \& Thilker D. 2010, Astronomy \& Astrophysics 511, A89\\
%\bibitem[\protect\citeauthoryear{Corbelli}{2010}]{Corbelli2010}
%Corbelli E., Lorenzoni S., Walterbos R., Braun R., Thilker D. 2010, Astronomy \& Astrophysics, 511, A89
%\bibitem[\protect\citeauthoryear{Casertano}{1983}]{Casertano1983}
%Casertano S. 1983, Mon. Not. R. Astron. Soc. 203, p.735\\
%\bibitem[\protect\citeauthoryear{Fich}{1989}]{Fich1989}
Feng J. Q., Gallo C. F. 2011, Research in Astron. Astrophys. Vol. 11, No. 12, p.1429\\
Fich M., Blitz L., Stark A. A. 1989, Astrophysical Journal 342, p.272\\
%\bibitem[\protect\citeauthoryear{Freeman}{1970}]{Freeman1970}
%Freeman K. C. 1970, Astrophysical Journal, 160, 811
%Lelli F., McGaugh S. S. \& Schombert S. 2016, The Astronomical Journal, 152, 157
%\bibitem{linshu1964} 
%Lin C. C. and Shu F. H. 1964,  Astrophysical Journal, 140, 646
%\bibitem{luna06} 
de Luna A., Bronfman L., Carrasco L. \& May J. 2006, The Astrophysical Journal 641, p.938\\
%\bibitem{jefrykaw82} Jaffrey A. \& Kawahara T. 1982,  Asymptotic Methods in Nonlinear Wave
%Theory (Pitman: Boston, MA)
%\bibitem{jefrytan64} Jaffrey A. \& Taniuti T. 1976, Non-linear Wave Propagation: with Applications
%to Physics and Magnetohydrodynamics (Academic Press: New York)
%\bibitem{mannheim12} 
Mannheim P. D. \&  O'Brien J. G. 2012, Phys. Rew. D 85, 12, 124020 \\
%\bibitem{milgrom1983} 
Milgrom M.1983, Astrophysical Journal 270, p.371\\
Moffat J. W. 2006, Journal of Cosmology Astroparticle Physics, \textbf{3}, 4\\
%\bibitem{ort1972} Oort J. H. 1972, \emph{Ann. NY Acad. Sci.}, \textbf{198}, 255
%\bibitem{rubin1970} 
%Rubin V. C.\& Ford Jr W. K. 1970, Astrophysical Journal, 159, 379
%\bibitem{ryder1998} 
%Ryder S. D., Zasov A. V., McIntyre V. J., Walsh W. \& Sil'chenko O. K. 1998, Mon. Not. R. Astron. Soc., 293, 411
%\bibitem{sanders1984} Sanders R. H. 1984, \emph{Astronomy \& Astrophysics}, \textbf{136}, L21
%\bibitem{santos2016} Santos-Santos I. M., Brook C. B., Stinson G., Di Cintio A., Wadsley J., DomÃ­nguez-Tenreiro R., GottlÃ¶ber S. \&  Yepes G. 2016, \emph{Mon. Not. R. Astron. Soc.}, \textbf{455}, 476
%\bibitem{slipher1917} Slipher, V. M. 1917, \emph{Lowell Observatory Bulletin}, \textbf{2}, VM1412
%\bibitem{sofue01} 
Sofue Y. \& Rubin V. 2001, Ann. Rev. Astron. \& Astrophys. 39, p.137\\
%\bibitem[\protect\citeauthoryear{Sofue}{2017}]{Sofue2018}
Sofue Y. 2018, PASJ 69 (1), 31, 1-35\\
%\bibitem[\protect\citeauthoryear{Sofue}{2017}]{Sofue2017}
%Sofue Y. 2017, PASJ, 70 (2), R1, 1-15
Springel V. 2005, Mon. Not. R. Astron. Soc. 364, p. 1105\\
Trachernach C., de Blok W. J. G., Walter F., Brinks E. \& Kennicutt Jr. R. C. 2008, The Astronomical Journal 136, 2720T\\
%\bibitem{verheijen1997} 
%Verheijen M. A. W. 1997, PhD Thesis, Rijksuniversiteit Groningen\\ 
%\bibitem{vukcevic14} 
Vukcevic M. 2014, Mon. Not. R. Astron. Soc. 441, p.565\\
%\bibitem[{vukcevic19}] 
Vukcevic M. 2019, Mon. Not. R. Astron. Soc. 484, p.3410\\
%\bibitem[\protect\citeauthoryear{Vukcevic}{2021}]{Vukcevic2021}
Vukcevic M. 2021, The Astronomical Journal 161, p.118\\
%\bibitem{watanabe1969} 
Watanabe T. 1969, Phys. Soc. Japan 27, p.1341\\

%[\protect\citeauthoryear{zwicky}{1937}]{zwicky37} 
%Zwicky F. 1937, Astrophysical Journal, 86, 217

%\nocite{*}% Show all bib entries - both cited and uncited; comment this line to view only cited bib entries;
%\bibliography{Wiley-ASNA}%

%\section*{Author Biography}
%(if applicable)

%\begin{biography}{\includegraphics[width=60pt,height=70pt,draft]{empty}}{\textbf{Author Name.} This is sample author biography text this is sample author biography text this is sample author biography text this is sample author biography text this is sample author biography text this is sample author biography text this is sample author biography text this is sample author biography text this is sample author biography text .}
%\end{biography}

\end{document}